# Efficient Identity Based Public Verifiable Signcryption Scheme

Prashant Kushwah[1] and Sunder Lal[2]

[1]Department of Mathematics and Statistics, Banasthali University, Rajasthan, India.
[2]Department of Mathematics, Dr. B. R. A. (Agra) University, UP, India.
Email:- [1]pra.ibs@gmail.com, [2]sunder_lal2@rediffmail.com

**Abstract:** Signcryption is a cryptographic primitive which performs encryption and signature in a single logical step. In conventional signcryption only receiver of the signcrypted text can verify the authenticity of the origin i.e. signature of the sender on the message after decrypting the cipher text. In public verifiable signcryption scheme anyone can verify the authenticity of the origin who can access the signcrypted text i.e. signature of the sender on the cipher text. Public verifiable signcryption scheme in which the receiver can convince a third party, by providing additional information other than his private key along with the signcryption is called third party verifiable signcryption schemes. In this paper we proposed an efficient identity based public verifiable signcryption scheme with third party verification and proved its security in the random oracle model.

**1. Introduction:** The main advantages of public key cryptography are encryption and digital signature, used to achieve confidentiality and authenticity of a message respectively. There are scenarios where both primitives are needed (for example secure e-mailing). Earlier signature-then-encryption approach was followed to achieve both primitives. However, this approach has high computational cost and communication overhead. In 1997, Zheng [17] proposed a novel cryptographic primitive "Signcryption" which achieves both confidentiality and authenticity in a single logical step with the cost significantly lower than 'signature-then-encryption' approach. In 2002, Beak et al. [1] first formalize and define security notions for signcryption via semantic security against adaptive chosen cipher text attack and existential unforgeability against adaptive chosen message attack. Many public key signcryption schemes have been proposed after [17]. Some of them are [2, 9, 10, 18].

Identity based cryptography was introduced by Shamir [15] in 1984. In the identity based cryptosystem public key of users are their identities (e.g. email address, PAN number etc.) and secret keys of users are created by a trusted third party called private key generator (PKG). First identity based signature scheme was given by Shamir [15] in 1984, but the first identity based encryption scheme was given by Boneh and Franklin [5] in 2001. The first identity based signcryption scheme was proposed by Malone Lee [12] in 2002. They also gave the security model for signcryption in identity based setting. Since then, many identity based signcryption schemes have been proposed in literature [3, 6, 7, 8, 11, 13]. Their main objective is to reduce the computational complexity and to design the more efficient identity based signcryption scheme.

In conventional signcryption the sender signs the message which is hide it the receiver's public key. Thus only the receiver can decrypt the message using his/her private key and can verify the authenticity of the cipher text. In the case when receiver wants to prove that indeed the sender has signed the message to a third party then he/she has to reveal his/her private key. In public verifiable signcryption scheme a third party who is unaware of the receiver's private key is able to verify whether a cipher text is valid or not. Public verifiable signcryption schemes have applications in filtering out the spam in a secure email system and private contract signing [14]. In third party verifiable signcryption schemes, a third party is able to verify the integrity and origin of the message using some additional information along with the signcryption provided by the receiver other than his/her private key. Recently in 2010, Selvi et al. [14] showed attacks on confidentiality and unforgeability of the Chow et al. [8] identity based signcryption scheme, which was the only identity based signcryption scheme that provides both public verifiability and third party verification. In [14]

Selvi et al. proposed a new identity based signcryption scheme with public verifiability and third party verification and suggested a modification in security notions by providing an additional oracle called third party verifiable (TP-Verify) oracle to the adversary. In this paper we propose an efficient identity based public verifiable signcryption scheme with third party verification and forward security. Also in the security model of [14] TP-Verify oracle does not provide any advantage to the adversary as it is already embedded in the IBPUSC oracle. Also in the proof of Theorem 1 [14], the simulation of TP-Verify oracle depends on the IBPUSC oracle. Thus we consider the security notions for identity based signcryption proposed in [6, 12].

This paper is organized as follows: In section 2, we define identity based signcryption scheme with public verifiability and third party verification and its security model. Section 3 contains the preliminaries for the proposed scheme. In section 4, we give the construction of IBPSC scheme and in section 5 we give the security results for our scheme under the security model defined in section 2. In section 6 we compare our scheme with the existing identity based signcryption schemes with similar properties. We conclude this paper in section 7.

## 2. Identity Based Public Verifiable Signcryption (IBPSC):

An **identity based public verifiable signcryption (IBPSC) scheme** consists of the following algorithms:

1. **Setup:** This algorithm takes input a security parameter k and outputs the system parameters **params** and a master secret key.
2. **Key Generation:** Given input params, master secret key and a user's identity $ID_U$, it outputs a partial private key $D_U$ corresponding to $ID_U$.
3. **IBPSC:** To send a message $m$ from a user $A$ to $B$, this algorithm takes input $(D_A, m, ID_A, ID_B)$ and outputs a $\sigma = IBPSC(D_A, m, ID_A, ID_B)$.
4. **IBPUSC:** This algorithm takes input $(\sigma, D_B, ID_B, ID_A)$ and outputs $m$ and $\phi$ if $\sigma$ is a valid signcryption of $m$ done by $A$ for $B$, otherwise outputs "invalid" if $\sigma$ is not valid.
5. **TP-Verify:** This algorithm takes input $(\phi, ID_A, ID_B)$ and outputs "Valid", if $\sigma$ is a valid signcryption of $m$ done by $A$ for $B$, otherwise "invalid", if $\sigma$ is not valid.

## Security model for IBPSC:

## 2.1. Message Confidentiality:

The notion of security with respect to confidentiality is indistinguishability of encryptions under adaptive chosen cipher text attack (IND-IBPSC-CCA2). For IBPSC this notion is captured by the following game played between challenger $\mathcal{C}$ and adversary $\mathcal{A}$.

## GAME 1 (IND-IBPSC-CCA2):

**Initialization:** $\mathcal{C}$ runs the setup algorithm on input a security parameter $k$, gives public parameters params to the adversary $\mathcal{A}$. $\mathcal{C}$ keeps the master key secret.

**Queries (Find Stage):** The adversary $\mathcal{A}$ makes the following queries adaptively.

➢ **Hash Queries:** $\mathcal{A}$ can request the hash values of any input and $\mathcal{C}$ responds with appropriate hash values.
➢ **Key generation Queries:** $\mathcal{A}$ submits an identity $ID_U$ and $\mathcal{C}$ computes the private key $D_U$ corresponding to $ID_U$ and returns to $\mathcal{A}$.

- ➤ **IBPSC Queries:** $\mathcal{A}$ submits two identities $ID_A$, $ID_B$ and a message $m$. Challenger $\mathcal{C}$ runs IBPSC algorithm with message $m$ and identities $ID_A$ and $ID_B$ and returns the output $\sigma$ to the adversary $\mathcal{A}$.

- ➤ **IBPUSC Queries:** $\mathcal{A}$ submits two identities $ID_A$, $ID_B$ along with $\sigma$ to the challenger $\mathcal{C}$. $\mathcal{C}$ runs the IBPUSC algorithm with input $\sigma$, $ID_A$ and $ID_B$ and returns the output $m$ and $\phi$ if $\sigma$ is a valid signcryption of $m$ done by $A$ for $B$, otherwise outputs "invalid" if $\sigma$ is not valid.

   No queries with $ID_A = ID_B$ is allowed.

**Challenge:** At the end of find stage, $\mathcal{A}$ submits two distinct messages $m_0$ and $m_1$ of equal length, a sender's identity $ID_A^*$ and a receiver's identity $ID_B^*$ on which $\mathcal{A}$ wishes to be challenged. The adversary $\mathcal{A}$ must have made no key generation query on $ID_B^*$. $\mathcal{C}$ picks randomly a bit $b \in \{0,1\}$, runs the IBPSC algorithm with message $m_b$ under $ID_A^*$ and $ID_B^*$ and returns the output $\sigma^*$ to the adversary $\mathcal{A}$.

**Queries (Guess stage):** $\mathcal{A}$ queries adaptively again as in the find stage. It is not allowed to extract the private key corresponding to $ID_B^*$ and also it is not allowed to make an IBPUSC query on $\sigma^*$ with sender $ID_A^*$ and receiver $ID_B^*$.

   Eventually, $\mathcal{A}$ outputs a bit $b'$ and wins the game if $b = b'$.

$\mathcal{A}$'s advantage is defined as $Adv_{\mathcal{A}}^{IND-IBPSC-CCA2} = 2\Pr[b = b'] - 1$.

**Definition 1:** An IBPSC scheme is said to IND-IBPSC-CCA2 secure if no polynomially bounded adversary $\mathcal{A}$ has non-negligible advantage of winning the above game.

   Note that the confidentiality game described above deals with the insider security since the adversary is given access to the private key of sender $ID_A^*$ in the challenge.

## 2.2. Cipher text unforgeability:

   The notion of security with respect to authenticity is existential unforgeability against chosen message attacks (EUF-IBPSC-CMA). For IBPSC this notion is captured by the following game played between challenger $\mathcal{C}$ and adversary $\mathcal{A}$.

**GAME 2 (EUF-IBPSC-CMA):**

**Initialization:** Same as in GAME 1.

**Queries:** The adversary $\mathcal{A}$ asks a polynomially bounded number of queries adaptively as in GAME 1.

**Forgery:** Finally, $\mathcal{A}$ produces a triplet $(ID_A^*, ID_B^*, \sigma^*)$ that was not obtained from IBPSC query during the game and for which private key of $ID_A^*$ was not exposed. The forger wins if $\sigma^*$ is valid signcrypted text from $ID_A^*$ to $ID_B^*$.

   The adversary $\mathcal{A}$'s advantage is its probability of winning the above game.

**Definition 3:** An IBPSC scheme is said to EUF-IBPSC-CMA secure if no polynomially bounded adversary $\mathcal{A}$ has non-negligible advantage of winning the above game.

Note that in the cipher text unforgeability game described above deals with the insider security since the adversary is given access to the private key of receiver $ID_B^*$ in the forgery.

## 3. Preliminaries:

Let $\mathbb{G}_1$ be an additive group and $\mathbb{G}_2$ be a multiplicative group both of the same prime order $p$. A function $e : \mathbb{G}_1 \times \mathbb{G}_1 \to \mathbb{G}_2$ is called a **bilinear pairing** if it satisfies the following properties:

1. $\forall P, Q \in \mathbb{G}_1, \forall a, b \in \mathbb{Z}_p^*, e(aP, bQ) = e(P, Q)^{ab}$
2. For any $\mathcal{O} \neq P \in \mathbb{G}_1$, there is $Q \in \mathbb{G}_1$, such that $e(P, Q) \neq 1$
3. There exists an efficient algorithm to compute $e(P, Q) \ \forall P, Q \in \mathbb{G}_1$.

Given a $(q+1)$ tuple $(P, aP, a^2P, ..., a^qP)$ to compute $e(P, P)^{1/a} \in \mathbb{G}_2$ is known as **q-Bilinear Diffie Hellman inversion problem (q-BDHIP)** [4].

Given $P, xP \in \mathbb{G}_1, h_1, ..., h_q \in_R \mathbb{Z}_p^*, \dfrac{1}{h_1 + x} P, ...., \dfrac{1}{h_q + x} P \in \mathbb{G}_1$ where $x \in_R \mathbb{Z}_p^*$ is unknown and $q$ is an integer, to compute $\dfrac{1}{h + x} P$ for some $h \in \mathbb{Z}_p^*$ but $h \notin \{h_1, ..., h_q\}$ is known as strong **q-Collision Attack Assumption problem (q-CAAP)** [16].

## 4. Proposed IBPSC Scheme:

**Setup:** Given a security parameter $1^k$, the PKG chooses two groups $\mathbb{G}_1$ and $\mathbb{G}_2$ of prime order $p > 2^k$, a random generator $P$ of $\mathbb{G}_1$, and a bilinear map $e : \mathbb{G}_1 \times \mathbb{G}_1 \to \mathbb{G}_2$, Computes $g = e(P, P)$, define hash functions as $H_1 : \{0,1\}^{k_3} \to \mathbb{Z}_p^*$, $H_2 : \{0,1\}^{n_1 + n_2 + 2k_1 + 2k_3} \to \{0,1\}^{n_2}$, $H_3 : \{0,1\}^{k_2 + 2k_1} \to \{0,1\}^{n_1 + n_2}$, $H_4 : \{0,1\}^{n_1 + n_2 + 2k_1 + 2k_3} \to \mathbb{G}_1$, where $k_1, k_2$ and $k_3$ denote the number of bits to represent elements of $\mathbb{G}_1$, $\mathbb{G}_2$ and identity respectively and $n_1$ is the message bit length and $n_2$ is the number of bits concatenated to message. PKG chooses random $s \in \mathbb{Z}_p^*$ as the master secret key and sets $P_{pub} = sP$. PKG publishes the system parameters as params $= \big\langle \mathbb{G}_1, \mathbb{G}_2, p, n_1, P, P_{pub},$ $e : \mathbb{G}_1 \times \mathbb{G}_1 \to \mathbb{G}_2, g, H_1, H_2, H_3, H_4 \big\rangle$.

**Key Generation:** Given a user $U$ with identity $ID_U$, the private key is computed by PKG as $D_U = (H_1(ID_U) + s)^{-1} P$. Also $Q_U = H_1(ID_U) P + P_{pub}$.

**IBPSC:** The sender $A$ for the receiver $B$

1. Chooses $r \in_R \mathbb{Z}_p^*$;
2. Computes
   i. $\alpha = g^{r^{-1}}$
   ii. $R = r^{-1} Q_B$ and $S = r Q_A$
   iii. $\gamma = H_2(m, \alpha, R, S, ID_A, ID_B)$
   iv. $c = m \| \gamma \oplus H_3(\alpha, R, S)$
   v. $H = H_4(c, R, S, ID_A, ID_B)$

vi. $T = rH + D_A$

3. Returns the signcrypted text $\sigma = (c, R, S, T)$.

**IBPUSC:** On receiving $\sigma$ from $A$, the user $B$

1. Computes $H' = H_4(c, R, S, ID_A, ID_B)$

2. If $e(T, Q_A) \neq e(H', S)g$ returns "invalid". Otherwise computes

   i. $\alpha' = e(R, D_B)$

   ii. $m' \| \gamma' = c \oplus H_3(\alpha', R, S)$

   iii. $\bar{\gamma} = H_2(m', \alpha', R, S, ID_A, ID_B)$

3. If $\bar{\gamma} \neq \gamma'$ returns "invalid". Otherwise returns $m'$ and $\phi = (m', \gamma', \alpha', \sigma)$.

**TP-Verify:** On receiving $\phi$ and $ID_A, ID_B$, Third party

1. Computes $m^* \| \gamma^* = c \oplus H_3(\alpha', R, S)$

2. Accept $\sigma$ and output valid iff $\gamma^* = H_2(m^*, \alpha', R, S, ID_A, ID_B)$ and $\gamma^* = \gamma'$. Otherwise outputs "invalid".

Note that in the proposed scheme $(R, S, T)$ can be seen as the signature of the sender $A$ on the cipher text $c$, which can be verified without the knowledge of receiver's private key. Thus the proposed identity based signcryption scheme achieves public verifiability. Also it is forward secure as the knowing of sender's private key does not help to decrypt the cipher text.

Consistency:

$$e(R, D_B) = e(r^{-1}Q_B, D_B) = e(H_1(ID_B)P + P_{pub}, D_B)^{r^{-1}}$$

$$= e((H_1(ID_B) + s)P, (H_1(ID_B) + s)^{-1}P)^{r^{-1}} = e(P, P)^{r^{-1}} = g^{r^{-1}}$$

$$e(T, Q_A) = e(rH + D_A, Q_A) = e(rH, Q_A)e(D_A, Q_A) = e(H, rQ_A)e(P, P) = e(H, S)g$$

## 5. Security Results:

**Theorem 1: (Message confidentiality)** Assume that an IND-IBPSC-CCA2 adversary $\mathcal{A}$ has an advantage $\varepsilon$ against the proposed IBPSC scheme when running in time $\tau$, asking $q_{h_i}$ queries to random oracle $H_i (i = 1, 2, 3, 4)$ and $q_e, q_u$ IBPSC queries, IBPUSC queries respectively. Then there is an algorithm $\mathcal{B}$ to solve the $q$-BDHIP for $q = q_{h_1}$ with probability

$$\varepsilon' > \frac{\varepsilon}{q_{h_1}(q_{h_3} + q_e)} \left(1 - q_u \left(\frac{q_{h_2}}{2^{n_2}} + \frac{1}{2^k}\right)\right) \left(1 - \frac{q_e(q_e + q_{h_4})}{2^k}\right)$$

within a time $\tau' < \tau + O(2q_u)\tau_p + O(q_{h_1}^2 + q_{h_4} + 3q_e)\tau_{multi} + O(q_e)\tau_{exp}$ where $\tau_{exp}$, $\tau_{multi}$ and $\tau_p$ are the time for an exponentiation in $\mathbb{G}_2$, multiplication in $\mathbb{G}_1$ and for a pairing computation.

**Proof:** Let $\mathcal{A}$ be an IND-IBPSC-CCA2 adversary against the proposed IBPSC scheme with advantage $\varepsilon$. We will show how adversary $\mathcal{A}$ is used to construct a simulator $\mathcal{B}$ that extract $e(P, P)^{1/a}$ on input $(P, aP, a^2P, ..., a^qP)$.

We will proceed similarly as in [3]. In the preparation phase, first $\mathcal{B}$ selects $\ell \in_R \{1,...,q_{h_1}\}$, elements $\lambda_\ell \in_R \mathbb{Z}_p^*$, $\mu_1, \mu_2,...,\mu_{\ell-1}, \mu_{\ell+1},...,\mu_q \in_R \mathbb{Z}_p^*$ and expand the polynomial $g(x) = \prod_{i=1, i \neq \ell}^{q} (x + \mu_i)$ to obtain the coefficients $c_1, c_2,....,c_{q-1} \in_R \mathbb{Z}_p^*$ such that $g(x) = \sum_{i=0}^{q-1} c_i x^i$. $\mathcal{B}$ also computes $\lambda_i = \lambda_\ell - \mu_i \in \mathbb{Z}_p^*$ for $i = 1,...,\ell-1, \ell+1,...,q$.

Now $\mathcal{B}$ sets $G = \sum_{i=0}^{q-1} c_i(a^i P) = g(a)P$ as a public generator of $\mathbb{G}_1$ and computes another element $U \in \mathbb{G}_1$ as $U = \sum_{i=1}^{q} c_{i-1}(a^i P) = aG$. Note that $\mathcal{B}$ does not know $a$. Further $\mathcal{B}$ computes

$$g_i(x) = \frac{g(x)}{(x + \mu_i)} = \sum_{i=0}^{q-2} d_i x^i$$

for $i = 1,...,\ell-1, \ell+1,...,q$ such that

$$\frac{1}{(a + \mu_i)}G = \frac{g(a)}{(a + \mu_i)} = g_i(a)P = \sum_{i=0}^{q-2} d_i(a^i P).$$

Thus $\mathcal{B}$ can compute $q - 1 = q_{h_1} - 1$ pairs $\left( \mu_i, S_i = \frac{1}{a + \mu_i}G \right)$ by the last term of the above equation. The system wide public key $P_{pub}$ is chosen as $P_{pub} = -U - \lambda_\ell G = (-a - \lambda_\ell)G$ with (unknown) private key $z = -a - \lambda_\ell \in \mathbb{Z}_p^*$. For all $i = 1,...,\ell-1, \ell+1,...,q$, $\mathcal{B}$ has $(\lambda_i, -S_i) = \left( \lambda_i, \frac{1}{\lambda_i + z}G \right)$.

Now simulator $\mathcal{B}$ start the interaction with $\mathcal{A}$ on input $\langle \mathbb{G}_1, \mathbb{G}_2, p, n_1, G, P_{pub}, e : \mathbb{G}_1 \times \mathbb{G}_1 \to \mathbb{G}_2, g, H_1, H_2, H_3, H_4 \rangle$ where $g = e(G,G)$ and $P_{pub} = zP$. $\mathcal{A}$ asks queries to $\mathcal{B}$ throughout the simulation. It is assumed that $H_1$ queries are distinct and any query involving the identity $ID$ comes after a $H_1$ query on $ID$. The target identity $ID_B^*$ is submitted to $H_1$ at some point of simulation. Also to maintain consistency in queries, $\mathcal{B}$ makes the lists $L_i$ for random oracle $H_i$ for $i = 1, 2, 3, 4$. $\mathcal{B}$ initializes a counter $\eta$ to 1 and start answering $\mathcal{A}$'s queries as follow:

- **$H_1$ queries:** it takes input an identity $ID$. $\mathcal{B}$ answers $\lambda_\eta$ to the $\eta^{\text{th}}$ one such query and increment $\eta$. $\mathcal{B}$ sets the identity $ID$ as $ID_\eta$ and store the tuple $(ID_\eta, \lambda_\eta)$ to $L_1$ list.

- **$H_2$ queries:** it takes input $(m, \alpha, R, S, ID_\zeta, ID_\eta)$. $\mathcal{B}$ checks the list $L_2$, it returns a previous value if it exists. Otherwise it chooses a random $h_2 \in_R \{0,1\}^{n_2}$ and returns this value as the answer. $\mathcal{B}$ stores the tuple $(m, \alpha, R, S, ID_\zeta, ID_\eta, h_2)$ in the $L_2$ list.

- **$H_3$ queries:** it takes input $(\alpha, R, S)$. $\mathcal{B}$ checks the list $L_3$, it returns a previous value if it exists. Otherwise it chooses a random $h_3 \in_R \{0,1\}^{n_1+n_2}$ and returns this value as the answer. $\mathcal{B}$ stores the tuple $(\alpha, R, S, h_3)$ in the $L_3$ list.

- **$H_4$ queries:** it takes input $(c, R, S, ID_\zeta, ID_\eta)$. $\mathcal{B}$ checks the list $L_4$, it returns a previous value if it exists. Otherwise it chooses a random $v \in_R \mathbb{Z}_p^*$ and returns $h_4 = vQ_\zeta \in \mathbb{G}_1$ as the answer. $\mathcal{B}$ stores the tuple $(c, R, S, ID_\zeta, ID_\eta, v, h_4)$ in the $L_4$ list.

- **Key generation queries:** it takes input an identity $ID_\eta$. $\mathcal{B}$ fails if $\eta = \ell$ and aborts the simulation. Otherwise it knows that $H_1(ID_\eta) = \lambda_\eta$ from $L_1$ list and returns $D_\eta = \frac{1}{\lambda_\eta + z}G$.

> **IBPSC queries:** it takes input a plaintext m and identities $(ID_\zeta, ID_\eta)$ where $\zeta, \eta \in \{1, ..., q_{h_i}\}$. If $\zeta \neq \ell$, $\mathcal{B}$ knows the private key of $ID_\zeta$ is $D_\zeta$ and can answer the query by following the specification of IBPSC algorithm. So we assume that $\zeta = \ell$, then $\mathcal{B}$ does the following:

    **i.** Chooses $r, t, v \in_R \mathbb{Z}_p^*$

    **ii.** Computes $\alpha = g^{r^{-1}}$, $R = r^{-1}Q_\eta$, $S = vP$, $T = tP$

    **iii.** Simulates $H_2(m, \alpha, R, S, ID_\ell, ID_\eta) = h_2$ and stores the tuple $(m, \alpha, R, S, ID_\ell, ID_\eta, h_2)$ in the $L_2$ list.

    **iv.** Simulates $H_3(\alpha, R, S) = h_3$ and stores the tuple $(\alpha, R, S, h_3)$ in $L_3$ list.

    **v.** Computes $c = m \parallel h_2 \oplus h_3$

    **vi.** Sets $H_4(c, R, S, ID_\ell, ID_\eta) = v^{-1}(tQ_\ell - P) \in \mathbb{G}_1$ and stores the tuple $(c, R, S, ID_\zeta, ID_\eta, \perp, h_4 = v^{-1}(tQ_\ell - P))$ in $L_4$ list.

    **vii.** Returns the signcryption $\sigma = (c, R, S, T)$

Also $\mathcal{B}$ fails if $H_4$ is already defined but this happens with a probability smaller than $(q_e + q_{h_4})/2^k$.

> **IBPUSC queries:** it takes input a signcrypted text $\sigma = (c, R, S, T)$ a sender's identity $ID_\zeta$ and a receiver's identity $ID_\eta$. If $ID_\eta \neq ID_\ell$ then $\mathcal{B}$ knows private key of $ID_\eta$ is $D_\eta$. $\mathcal{B}$ runs the IBPUSC algorithm normally and returns the output to $\mathcal{A}$. If $ID_\eta = ID_\ell$, then $\mathcal{B}$ searches the $L_4$ list for the entry $(c, R, S, ID_\zeta, ID_\eta, v, h_4)$. If such an entry does not exist then $\mathcal{B}$ returns invalid, otherwise it computes $e(T, Q_\zeta)$ and $e(h_4, S)g$ where $g = e(G, G)$. If $e(T, Q_\zeta) \neq e(h_4, S)g$, $\mathcal{B}$ returns invalid, otherwise for each tuple $(\alpha_i, R, S, h_{3,i})$ in $L_3$ list $\mathcal{B}$ computes $m_i \parallel \gamma_i = c \oplus h_{3,i}$ and searches the $L_2$ list for the tuple $(m_i, \alpha_i, R, S, ID_\zeta, ID_\eta, h_{2,i})$. If $\gamma_i \neq h_{2,i}$ for any $i$, $\mathcal{B}$ returns "invalid", otherwise returns $\phi = (m_i, \gamma_i, \alpha_i, \sigma)$.

Across the whole game the probability to incorrectly reject the signcrypted text at some moment of the simulation is bounded by $q_u\left(\dfrac{q_{h_2}}{2^{n_2}} + \dfrac{1}{2^k}\right)$.

At the end of challenge phase, $\mathcal{A}$ outputs two messages $m_0, m_1$ and identities $ID_A^*$, $ID_B^*$ such that she has not made Key generation query on $ID_B^*$. If $ID_B^* \neq ID_\ell$, $\mathcal{B}$ aborts the simulation. Otherwise it picks $\xi \in_R \mathbb{Z}_p^*$, $c \in_R \{0,1\}^{n_1+n_2}$ and $S, T \in_R \mathbb{G}_1$ to return the challenge $\sigma^* = (c, R, S, T)$ where $R = -\xi G \in \mathbb{G}_1$. If we define $\delta^{-1} = \xi/a$ and since $z = -a - \lambda_\ell$, we can check that

$$R = -\xi G = -a\delta^{-1}G = (\lambda_\ell + z)\delta^{-1}G = \delta^{-1}\lambda_\ell G + \delta^{-1}P_{pub}$$

$\mathcal{A}$ cannot recognize that $\sigma^*$ is not a valid signcrypted text unless she queries $H_3$ with input $e(G,G)^{\delta^{-1}}$. Also in the guess stage, her view is simulated as before and her eventual output is ignored. Standard arguments can show that a successful $\mathcal{A}$ is very likely to query $H_3$ with input $e(G,G)^{\delta^{-1}}$ if the simulation is indistinguishable from a real attack environment.

To produce a result, $\mathcal{B}$ fetches a random record from $L_3$ list. As $L_3$ contains no more than $(q_{h_3} + q_e)$ records by construction thus with probability $\dfrac{1}{(q_{h_3} + q_e)}$, $\mathcal{B}$ chooses the record which will contain the right element $e(G,G)^{\delta^{-1}} = e(P,P)^{g(a)^2 \xi / a}$ where $G = g(a)P$.

The q-BDHIP solution can be extracted as follows, if $\omega^* = e(P,P)^{1/a}$ then

$$e(G,G)^{1/a} = (\omega^*)^{c_0^2} e(\textstyle\sum_{i=0}^{q-2} c_{i+1}(a^i P), c_0 P) e(G, \textstyle\sum_{j=0}^{q-2} c_{j+1}(a^j P))$$

In an analysis of $\mathcal{B}$'s advantage, following events will cause $\mathcal{B}$ to abort the simulation:

$E_1$: $\mathcal{A}$ does not choose to be challenge on $ID_\ell$

$E_2$: a Key generation query is made on $ID_\ell$

$E_3$: $\mathcal{B}$ aborts in IBPSC query because of a collision on $H_4$

$E_4$: $\mathcal{B}$ rejects a valid signcrypted text at some point of the game

We clearly have probability $\Pr[\neg E_1] = 1/q_{h_1}$ and we know that $\neg E_1$ implies $\neg E_2$. Also $\Pr[E_3] \le q_e(q_e + q_{h_4})/2^k$ and $\Pr[E_4] \le q_u\left(\dfrac{q_{h_2}}{2^{n_2}} + \dfrac{1}{2^k}\right)$. Thus we find that

$$\Pr[\neg E_1 \wedge \neg E_3 \wedge \neg E_4] \ge \frac{1}{q_{h_1}}\left(1 - q_u\left(\frac{q_{h_2}}{2^{n_2}} + \frac{1}{2^k}\right)\right)\left(1 - \frac{q_e(q_e + q_{h_4})}{2^k}\right)$$

Also the probability that $\mathcal{B}$ select the correct record from $L_4$ list is $\dfrac{1}{(q_{h_3} + q_e)}$. Therefore advantage of $\mathcal{B}$ is

$$\varepsilon' > \frac{\varepsilon}{q_{h_1}(q_{h_3} + q_e)}\left(1 - q_u\left(\frac{q_{h_2}}{2^{n_2}} + \frac{1}{2^k}\right)\right)\left(1 - \frac{q_e(q_e + q_{h_4})}{2^k}\right).$$

The time bound is obtained as there are $O(q_{h_1}^2 + q_4 + 3q_e)$ multiplications in preparation phase, $O(2q_u)$ pairing computations and $O(q_e)$ exponentiations in $\mathbb{G}_2$.

**Theorem 2 (Cipher text Unforgeability):** Assume that an EUF-IBPSC-CMA adversary $\mathcal{A}$ has an advantage $\varepsilon$ against the proposed IBPSC scheme when running in time $\tau$, asking $q_{h_i}$ queries to random oracle $H_i (i = 1, 2, 3, 4)$ and $q_e, q_u$ IBPSC queries, IBPUSC queries respectively. Then there is an algorithm $\mathcal{B}$ to solve the q-CAA problem for $q = q_{h_1} - 1$ with probability

$$\varepsilon' > \left(\varepsilon - \frac{1}{2^k}\right)\frac{1}{q_{h_1}(q_{h_3} + q_e)}\left(1 - q_u\left(\frac{q_{h_2}}{2^{n_2}} + \frac{1}{2^k}\right)\right)\left(1 - \frac{q_e(q_e + q_{h_4})}{2^k}\right)$$

within a time $\tau' < \tau + O(2(q_u + 1)\tau_p + O(q_{h_1}^2 + q_{h_4} + 3q_e)\tau_{multi} + O(q_e)\tau_{exp}$ where $\tau_{exp}$, $\tau_{multi}$ and $\tau_p$ are same as in theorem 1.

**Proof:** Let $\mathcal{A}$ be the EUF-IBPSC-CMA adversary against the proposed IBPSC scheme with advantage $\varepsilon$. We will show how adversary $\mathcal{A}$ is used to construct a simulator $\mathcal{B}$ that solve the q-CAA problem given input $P, sP \in \mathbb{G}_1$ (s is unknown), $\lambda_1, ..., \lambda_q \in_R \mathbb{Z}_p^*$ and $\frac{1}{\lambda_1 + s}P, ..., \frac{1}{\lambda_q + s}P \in \mathbb{G}_1$ i.e. $\mathcal{B}$ outputs a pair $(\lambda, \frac{1}{\lambda + s}P)$ for $\lambda \notin \{\lambda_1, ..., \lambda_q\}$ where $q = q_{h_1} - 1$.

Simulator $\mathcal{B}$ starts interaction with $\mathcal{A}$ on input $\langle \mathbb{G}_1, \mathbb{G}_2, p, n_1, P, P_{pub}, e : \mathbb{G}_1 \times \mathbb{G}_1 \rightarrow \mathbb{G}_2,$ $g, H_1, H_2, H_3, H_4 \rangle$ where $P_{pub} = sP$ and $g = e(P, P)$. $\mathcal{A}$ asks queries throughout the simulation. It is assumed that $H_1$ queries are distinct and any query involving the identity $ID$ comes after a $H_1$ query on $ID$. Also $\mathcal{B}$ makes list $L_i$ for random oracle $H_i, i = 1, 2, 3, 4$ to maintain consistency in queries as in theorem 1. $\mathcal{B}$ randomly picks $\ell \in_R \{1, ..., q_{h_1}\}$ and start answering $\mathcal{A}$'s queries as follows:

**$H_1$ queries:** It takes input an identity $ID$. $\mathcal{B}$ answers $\lambda_\eta \in \{\lambda_1, ..., \lambda_q\}$ to $\eta^{th}$ one such query and increment $\eta$. $\mathcal{B}$ sets the identity as $ID_\eta$ and store the tuple $(ID_\eta, \lambda_\eta, \frac{1}{\lambda_\eta + s}P)$ in $L_1$ list. If $\eta = \ell$, $\mathcal{B}$ returns a value $\lambda_\ell \in_R \mathbb{Z}_p^*$ such that $\lambda_\ell \notin \{\lambda_1, ..., \lambda_q\}$. $\mathcal{B}$ sets the identity as $ID_\ell$ and store the tuple $(ID_\ell, \lambda_\ell, \perp)$ in $L_1$ list.

**$H_2, H_3, H_4$ queries:** $\mathcal{B}$ answers these queries similarly as in theorem 1.

**Key Generation queries:** It takes input an identity $ID_\eta$. $\mathcal{B}$ fails if $\eta = \ell$ and aborts the simulation. Otherwise $\mathcal{B}$ checks the list $L_1$ to find the entry $(ID_\eta, \lambda_\eta, \frac{1}{\lambda_\eta + s}P)$ and returns $D_\eta = \frac{1}{\lambda_\eta + s}P$ as corresponding private key.

**IBPSC, IBPUSC queries:** $\mathcal{B}$ answers these queries similarly as in theorem 1.

At the end of the game, the forger $\mathcal{A}$ halts and outputs a signcrypted text $\sigma^* = (c^*, R^*, S^*, T^*)$ and two identities $ID_A^*$ and $ID_B^*$ such that $\sigma^*$ is not the output of IBPSC oracle with a sender's identity $ID_A^*$ and a receiver's identity $ID_B^*$. If $ID_A^* \neq ID_\ell$, $\mathcal{B}$ aborts the simulation. Otherwise $\mathcal{B}$ searches the $L_4$ list for the tuple $(c^*, R^*, S^*, ID_\ell, ID_B^*, v^*, h_4^* = v^* Q_\ell)$ such that

$$e(T^*, Q_\ell) = e(h_4^*, S^*)e(P, P)$$

with $h_4^* = v^* Q_\ell$ for some known elements $v^* \in \mathbb{Z}_p^*$. Then we have

$$e(T^* - v^* S^*, Q_\ell) = e(P, P)$$
$$e(T^* - v^* S^*, (\lambda_\ell + s)P) = e(P, P)$$

Thus $(\lambda_\ell + s)(T^* - v^* S^*) = P$. Hence $\mathcal{B}$ can successfully compute $\frac{1}{\lambda_\ell + s}P = T^* - v^* S^*$ and output the pair $(\lambda_\ell, \frac{1}{\lambda_\ell + s}P)$ for $\lambda_\ell \notin \{\lambda_1, ..., \lambda_q\}$ as a solution of k-CAA problem in $\mathbb{G}_1$.

The probability for $\mathcal{A}$ to output a valid forgery $\sigma^* = (c^*, R^*, S^*, T^*)$ without asking the corresponding $H_4(c^*, R^*, S^*, ID_\ell, ID_B^*)$ query is at most $1/2^k$.

Further the following events will cause $\mathcal{B}$ to abort the simulation:

$E_1$: $\mathcal{A}$ does not choose to be challenge on $ID_\ell$

$E_2$: a Key generation query is made on $ID_\ell$

$E_3$: $\mathcal{B}$ aborts in IBPSC query because of a collision on $H_4$

$E_4$: $\mathcal{B}$ rejects a valid signcrypted text at some point of the game

The analysis of the $\mathcal{B}$'s advantage is similar as in theorem 1. Therefore advantage of $\mathcal{B}$ is

$$\varepsilon' > \left( \varepsilon - \frac{1}{2^k} \right) \frac{1}{q_{h_1}(q_{h_3} + q_e)} \left( 1 - q_u \left( \frac{q_{h_2}}{2^{n_2}} + \frac{1}{2^k} \right) \right) \left( 1 - \frac{q_e(q_e + q_{h_4})}{2^k} \right).$$

The time bound is obtained as there are $O(q_{h_1}^2 + q_4 + 3q_e)$ multiplications in preparation phase, $O(2(q_u + 1))$ pairing computations and $O(q_e)$ exponentiations in $\mathbb{G}_2$.

## 6. Efficiency and Comparison:

Chow et al. [8] scheme was the only identity based signcryption scheme that provides both public verifiability and third party verification and supported by security proof. Recently, Selvi et al. [14] showed attacks on confidentiality and unforgeability of [8] and proposed a new identity based signcryption scheme with public verifiability and third party verification. Thus in the following table we compare our scheme with [14]. Clearly in the proposed scheme only three pairing computations are needed in the unsigncryption phase and no pairing calculation is needed in signcryption phase.

| Scheme | Signcryption | | | Unsigncryption | | |
|---|---|---|---|---|---|---|
| | mul in $\mathbb{G}_1$ | exps in $\mathbb{G}_2$ | e cps | mul in $\mathbb{G}_1$ | exps in $\mathbb{G}_2$ | e cps |
| Selvi et al. [14] | 2 | 1 | 1 | 0 | 0 | 4 |
| Proposed IBPSC | 3 | 1 | 0 | 0 | 0 | 3 |

Table 1

## 7. Conclusion:
In this paper we proposed an efficient identity based signcryption scheme with public verifiability and third party verification. In the proposed scheme only three pairing computations are needed in the unsigncryption phase and no pairing calculation is needed in signcryption phase. We compare our scheme with the existing identity based signcryption schemes with similar properties. We also gave the proofs of security based on q-BDHIP and q-CAA problem in the random oracle model.